\RequirePackage{fix-cm}
\documentclass[smallextended]{svjour3}       
\smartqed  
\usepackage{graphicx}
  \journalname{Special Issue on the foundations of astrophysics and cosmology}
\begin{document}

\title{Hubble law :  measure and interpretation }



\author{Paturel Georges \and 
Teerikorpi Pekka \and 
Baryshev Yurij }

\authorrunning{Paturel, Teerikorpi, Baryshev} 

\institute{G. Paturel \at
Retired from Universite Claude-Bernard, Observatoire de Lyon, 69230 Saint-Genis Laval, France \\
 Tel.: +334 78560474\\
 \email{   g.patu@orange.fr   }           
  \and
 P.  Teerikorpi \at
 Tuorla Observatory, Department of Physics and Astronomy, University of Turku, 21500 Piikki\"{o}, Finland
 \email{  pekkatee@utu.fi   } 
 \and
Y. Baryshev  \at
 Department of Astronomy, St. Petersburg State University, Staryj Peterhoff, 198504, St.Petersburg, Russia
 \email{  yubaryshev@mail.ru  } 
}

\date{Received: date / Accepted: date}

\maketitle

\begin{abstract}
We have had the chance to live through a fascinating revolution in measuring the fundamental empirical cosmological Hubble law. The key progress is analysed :  1) improvement of observational means (ground-based radio and optical observations, space missions) ;  2) understanding of the biases that affect both distant and local determinations of the Hubble constant;  3) new theoretical and observational results. These circumstances encourage us to take a critical look at some facts and ideas related to the cosmological red-shift. This is important because we are probably on the eve of a new understanding of our Universe, heralded by the need to interpret some cosmological key observations in terms of unknown processes and substances. 
\end{abstract}

\section{Introduction}
\label{intro}
This paper is a short review concerning the study of  the Hubble law. The purpose is to give an overview of  the evolution of cosmology with a presentation as simple as possible.

In the present section,  we give a brief description of the first steps in both conceptual and technical improvements that has led us to 
the present situation.
In section 2, we explain the biases that plagued for decades the determination of the so-called Hubble constant. In section 3, we summarize an analysis of the remaining biases that may cause some discordant results in current determinations of the Hubble constant. In section 4,  we discuss some intriguing questions connected to the Hubble law, to show that some progresses are still needed. Finally, in section 5, we conclude by trying to see how cosmology could further evolve.

Before 1925, Slipher was measuring\cite{vesto} radial velocities of nebulae. He noted, without interpretation, that the radial velocity of  some spiral nebulae increased with distance. Hubble proved that some of these "stellar systems" are indeed galaxies, similar to our Milky Way.  Lema\^{\i}tre\cite{lemaitre}, Lundmark\cite{lundmark} and Hubble  \cite{edwin} made an analysis of the relation between the radial velocity $V$\footnote{$V$ may be replaced by $c\,z$, where $c$ is the velocity of light and $z$ the so-called "red-shift".}
and the distance $r$  (the luminosity distance) measured from apparent magnitude :
\begin{equation}
V= H_0 \  r ,
\label{hubl}
\end{equation} 
where $H_0$  is the Hubble constant. This is the first important parameter in cosmology. The linearity of this relation is now well established for the Local Universe  ($V<30000$km/s). The main observational problem is the accuracy of the determination of $H_0$.

For a given galaxy, $H_0$ is obtained by dividing its radial velocity (in km/s) by its distance (in Mpc). The radial velocity is measured with a high accuracy (though there may be some unknown peculiar velocity), but not the distance. It was especially the sources of error in distance measurements, which, when corrected, caused the decrease of the measured values of $H_0$ during the last century. From the preliminary values of $500 \pm 50$ , quickly corrected to $200 \pm 50$ (km/s)/Mpc  by Baade, a continuous evolution took place. By the 1970s, the value of $H_0$ was reduced to 50-100 (km/s)/Mpc. A well-known debate \cite{deba1,deba2} started around these numbers. The leaders were Sandage and de Vaucouleurs. The Hubble constant was roughly 50 (km/s)/Mpc for the first team and 100 (km/s)/Mpc  for the second one.

Before 1970, only a few methods existed to measure distances to remote galaxies. They were often based on very poor correlations between the absolute magnitude and some observable parameters (brightness, morphological appearance, colour).

After 1970, with the development of electronics and computers, photographic plates, photomultipliers (PM), brightness-amplifiers (used in conjunction with ordinary photographic plates) and electronic cameras (used with thick photographic plates) were progressively replaced by  high performance Charge-Coupled-Devices (CCD). A telescopic exposure on a galaxy dropped from 2 hours to one minute or less. Further, the image was directly stored in the computer, ready to be analysed. The photometry of galaxies became much better than before. 
	The historical discovery of galactic radio emission was made by Karl Jansky in 1933, but the real development of radio-astronomy started around 1960-1970, due to the technological improvements, that quickly made this new wavelength window a major way to obtain accurate radial velocities (better than 20 km/s to be compared to 100 km/s by optical means).  Radio-astronomy created a revolution in the determination of extragalactic distances. 
At the same time (1980), multi-object spectrograph's (e.g. the pioneering  fibre-optic FLAIR Australian UK Schmidt telescope), allowed the measurement of hundreds of galaxies at once.
	Later (1990), we entered the new epoch of spatial missions with HIPPARCOS and the Hubble Space Telescope (HST), two fundamental instruments for the distance scale, and COBE, WMAP and PLANCK for the analysis of the Cosmic Microwave Background radiation from the era of the first atoms. At the same time ground based wide or deep digital surveys were undertaken in different wavelengths, from ultra-violet to infra-red (e.g. DENIS, TWOMASS, SLOAN).

\section{Bias in extragalactic measurements}
\label{sec1}

\subsection{The Tully-Fisher relation}
\label{sec2.1}
Very early after the first extragalactic radio-astronomy measurements, it was understood  \cite{Ref6,Ref7} that the width of the 21-cm line of neutral hydrogen (HI) contained information on the internal rotation energy of a galaxy. Through the virial theorem, it was expected that this HI-width, eventually combined with others parameters, might  produce a good correlation with the total mass and thus, with the absolute magnitude, which is the key to get accurate distance-moduli $\mu$ from the apparent magnitude $m$ :
\begin{equation}
\mu = 5  \ \log r_{Mpc}+ 25 = m - M ,
\label{mu}
\end{equation}
where $r$ is the radial  luminosity-distance in Mpc (megaparsec), $ m$ is the apparent magnitude corrected for dust extinction and inclination effects, and $M$ the absolute magnitude, obtained from the correlation with the HI-width and (eventually) additional parameters. 
It is to be noted that relation \ref{mu} can produce {\it{relative}}, unbiased distances  moduli by replacing $r$ by the radial velocity properly corrected for known peculiar motions. This assumes that the Hubble law is linear in the explored domain.This will be used to calculate {\it{relative, absolute magnitude}} from $V$ and $m$, as in Fig. \ref{spaen}.

The famous paper written by Tully \& Fisher \cite{tf} showed that the HI-width (expressed in a logarithmic scale) is linearly correlated with the absolute magnitude, without additional parameters: 
\begin{equation}
M = a \ \log W_{HI} + b .
\label{tfr}
\end{equation}

In Eq. \ref{tfr}  (the TF relation)  $ a$ and $b$ are two empirical constants. $W_{HI}$ is the 21cm line width (the unit of  $W_{HI}$ is arbitrarily chosen, but generally it is expressed in km/s, in order to normalize the numerical values of $a$ and $b$ parameters), corrected for some non-intrinsic effects (e.g., the inclination). 

 Indeed, the dispersion of published values of $H_0$ has been reduced, but not sufficiently for the debate to be considered as over.
Apparently, the only remaining problem seemed to be the primary calibration expected from the spatial mission  HIPPARCOS. However, the subject proved to be more complex, as we explain below.

\subsection{The Malmquist bias}
\label{sec2.2}

Let us explain the bias in an easy way by using "Sosie galaxies"\footnote{ Sosie is a French word for twins not genetically linked.}. 
Imagine we select galaxies having very nearly the same observed parameters: $W_{HI}$, morphological type, inclination. Further, let us assume that one of these galaxies is a {\it{calibrator}} with its distance known from a primary calibration. 
According to the TF relation (Eq. \ref{tfr}), the selected galaxies must all have the same absolute magnitude independently of the slope $a$ and of any inclination and morphological type dependencies. 
From Eq. \ref{mu} and \ref{tfr}, the distance modulus of each selected galaxy is simply:

\begin{equation}
\mu = \mu(calibrator) + m - m(calibrator).
\label{mu2}
\end{equation}  

One of us (GP) used this method and found  $H_0  \simeq 90$ (km/s)/Mpc. This is so straightforward that he could not imagine the result to be incorrect.  One of us (PT), a specialist of the different kinds of bias \cite{bias}, helped understand the hidden bias that he called the Malmquist bias of the 2nd kind \cite{2kind}. The origin of the bias has its roots in both the cosmic dispersion of intrinsic properties and the observational completeness limit. 

\begin{figure}
\resizebox{0.6\hsize}{!}{ \includegraphics{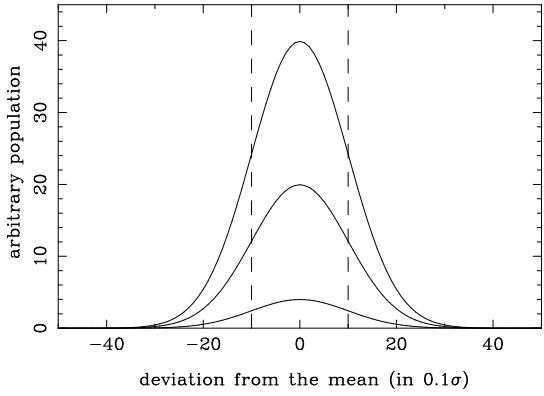}}
\caption{ Three Gaussian curves describing how a measured parameter (e.g. $W(HI)$ for a population of galaxies) is spread around the mean value, assuming that the mean is zero and that the standard deviation is constant, but with increasing size of the sample (from bottom to the top). The probability to find points far from the mean increases with the size of the sample (see  \cite{jnt}).}
\label{gauss}    
\end{figure}

Let us explain first the cosmic dispersion :  For a same rotation velocity (i.e. a same HI-width) two galaxies may have different absolute magnitudes. Indeed, a galaxy is a complex object the properties of which result from the full history of its evolution (e.g., the composition of matter at the time of its formation, interactions with its neighbourhood, etc.). This creates a cosmic dispersion, in addition to that of measurements. The larger the number of objects, the larger the width of the distribution, as shown in Fig. \ref{gauss}.

\begin{figure}
  \resizebox{0.6\hsize}{!}{\includegraphics{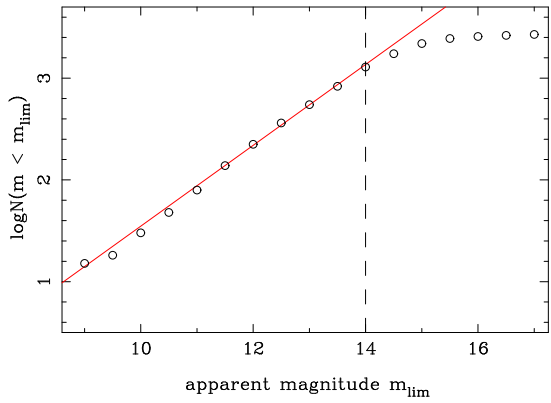}}
\caption{A plot of $logN(m<m_{lim})$ {\it{vs.}} apparent magnitude $m_{lim}$ gives the completeness limit $m_{lim}$ where the slope bends down, when the number $N$ of objects does not grow any more as the volume (see text). For a homogeneous distribution of galaxies the linear slope is $0.6$.}
\label{complet}    
\end{figure}

Then, let us determine the completeness limit of a galaxy sample by using the plot of $logN$  ($N$ is the number of objects) versus the apparent magnitude (see Fig.  \ref{complet}). The limiting magnitude (vertical dashed line) is given by the point where the curve starts to bend down, when the number of objects does not grow any more as the volume, because some objects are progressively too faint to be observed\footnote{Some objects may be excluded from the sample because of an incomplete set of data, as it will be seen for Cepheids.}.
 
\begin{figure}
   \resizebox{0.7\hsize}{!}{\includegraphics{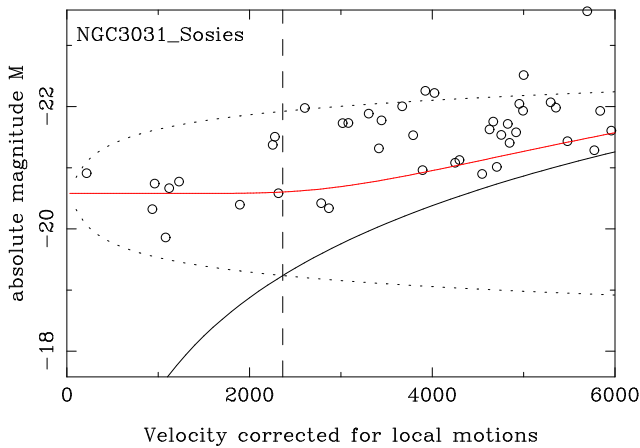}}
\caption{ Example of a  Spaenhauer diagram for Sosie galaxies of NGC 3031. The galaxies are selected to have all the same absolute magnitude (see text). The dispersion increases with distance (doted curves) estimated by radial velocity. Galaxies below the completeness limit (black curve) are missing. The vertical dashed line shows the limit of the completeness. The fit of data (red curve) gives unbiased mean on the left of  the dashed line. On the right of it, the red curve shows the biased mean, growing with distance. }
\label{spaen}    
\end{figure}

The Spaenhauer diagram \cite{spaenh} will make the bias visible.
This is the plot of the {\it{relative, absolute magnitude}} (from Eq.\ref{mu} as explained before) versus the radial velocity.
It shows  (Fig. \ref{spaen}) the combination of both effects (dispersion and completeness), for Sosie galaxies assumed to have the same absolute magnitude.  Beyond a certain distance (vertical line in Fig. \ref{spaen}), the estimate of the mean absolute magnitude changes. It becomes smaller (upper part of the plot), due to the magnitude limit, cutting away galaxies from the fainter half of the distribution of absolute magnitudes at large distances. Two features appear as a characteristic of bias : 1) the fit of the points (red curve) is not linear and its scatter is smaller at large distances.

In 1986, our team took this bias into account \cite{Ref14,Ref15} and found that the Hubble constant is not 90(km/s)/Mpc, but rather $63 \pm 3$(km/s)/Mpc or $72 \pm 3$(km/s)/Mpc, depending on the primary calibration (from Sandage \& Tammann or de  Vaucouleurs,  respectively).

One problem remains : the primary calibration. 
In 1997, the results of the HIPPARCOS astrometric satellite were published, from which good parallaxes were  expected for a large sample of Cepheids. But it was more difficult than thought. Indeed, the measurements of parallaxes are made difficult because of another bias (the Lutz \& Kelker bias \cite{lutz}). Finally, among a sample of 36 Galactic Cepheids within $1$kpc, only two have an acceptable parallax ($\delta$ Ceph and $\alpha$ UMi - the last one being excluded as an overtone variable star). Feast and Catchpole (\cite{feast}) made the best use of  HIPPARCOS data with a larger statistical study of 233 Galactic Cepheids, of which  only 26 have a significant contribution. Following the rule of thumb given by Lutz \& Kelker: {\it{ "You should not trust a distance greater than about 0.175 time the theoretical limit of your device"}}. So, the primary calibration has still to be improved.

In 2001, the Hubble Space Telescope provided astronomers with unprecedented data by measuring extragalactic Cepheids, up to the Virgo cluster, during a HST-KeyProject  (HSTKP \cite{Ref19}). Their distances through the Period-Luminosity (hereafter PL) relation are calibrated with the Large Magellanic Cloud.

 Although the mathematical forms of the PL relation and of the TF relation are very similar, the PL relation was assumed to be free of any bias, because of its relatively small dispersion. Thus, the global value of the Hubble constant derived from the HSTKP measurement ($H_0=72 \pm 8 $ (km/s)/Mpc) confirmed the reality of the Malmquist bias, as emphasized  by one of us (PT) since 1984 \cite{pt84}. 
It seemed that the dilemma between the high and low $H_0$ was resolved with a median value, even if some improvements are expected from future primary calibrations. Great hopes are still put on GAIA, the next generation of astrometric satellites, launched in December 2013. Indeed, its precision is very impressive: 40 times deeper than HIPPARCOS, parallaxes with a precision of 7 $\mu$as  (micro arcsecond) for magnitude of 10 and 0.1 mas (milli arcsecond) for magnitude 20. Combined with good long distance criteria, GAIA (e.g. Beaton et al. \cite{beaton}) may lead to a significant improvement on $H_0$.

\subsection{Recent improvements}
\label{sec2.3}

During the process of clarifying the debate about the value of $H_0$, some major discoveries were made in cosmology. The most important came from the possibility to detect and measure many Super Novae  explosions (SN), in particular Type Ia SN.  This unpredictable and relatively short phenomenon is very intense and can be detected at extremely large distances, in a region not yet explored for distance determination.  The SNIa method  (up to $z=0.1$) gave $H_0=63.1\pm3.4$(internal)$\pm2.9$(external) (km/s)/Mpc \cite{hamuy}.   Two teams \cite{perl,riess} discovered that the Hubble constant seemed to be smaller in the past, as if the expansion of the Universe was accelerating. It seemed compulsory to re-introduce the cosmological constant $\Lambda$, initially added by Einstein to force the Universe to be static. Now it is used to explain the acceleration of the expansion as described by the $\Lambda$CDM model with cosmological constant and cold dark matter (CDM).  For $z$ from $0.1$ to $1.0$ the value of $H$ seems to decrease.

In 2013 and 2016, the WMAP \cite{wmap} and PLANCK \cite{planck} satellites, devoted to the study of the Cosmic Microwave Background  (CMB) radiation, previously discovered  in 1964 by Penzias \& Wilson, obtain all combined parameters of the standard $\Lambda$CDM model. The Hubble constant $H_0$  (derived from CMBR observations related to $ z \simeq 1200)$  is $69.32\pm0.80$ and $67.8\pm 0.9$, for WMAP and PLANCK, respectively.

On the other hand, several methods were proposed to improve the quality of distances, in particular for the primary calibration:

1)  Barnes-Evans method  applied  to Cepheids \cite{gieren}. It is a quasi-geometrical method that requires difficult measurements but which is only  slightly model dependent. 

2) Parallaxes measurements with the HST \cite{benedict} reach a precision of 0.2 milliarsec (mas), i.e. better than the HIPPARCOS astrometric satellite that  reached a precision of about 1 mas.  

3) In addition some occasional observations (binary Cepheid stars, Eclipsing stars, kinematics of  maser sources in distant galaxies) permit a significant improvement  of the calibration. Let us describe briefly the very promising maser method in the NGC4258 case. The Keplerian motion of maser sources around a dense object in the host galaxy NGC4258 observed by Herrnstein et al. \cite{4258}  leads, through tangential and radial velocity measurements, to a geometrical distance modulus of $29.28\pm0.08$   

4) Further, some Cepheids were observed in galaxies in which SNIa events were observed \cite{hoffmann}. This makes the link between short and long distance observations. 

The most recent paper which exploited all resources available today (in 2016) has been published by Riess et al. (\cite{riess2}). This paper gives a Hubble value with an incredibly small uncertainty: $H_0=73.02 \pm 1.79$ (km/s)/Mpc. They suggested  that the difference from the global PLANCK value (67.8) could be significant, perhaps indicating some new cosmological effect. However, it is possible that the Cepheid bias discussed below and not studied in \cite{riess2}, could well cause such a shift of a few units upwards.

Other important work has been done to improve the local description of the cosmic flow of galaxies, and thus, the observed  local value of $H_0$, that must be corrected for peculiar motions. This has been done through a wide collaboration conducted by H. Courtois, B. Tully and collaborators \cite{laniakea}. The initial project did not aim at a precise estimation of the Hubble constant. Nevertheless, the calculation being made with more than 8315 distance determinations, from different methods, it leads to a good description of peculiar velocities and thus, to the resulting Hubble flow. The value for $H_0$ reported in the work is $74.4 \pm 3.0$ (km/s)/Mpc. It  is consistent with the recent observed values mentioned above. However, it is not clear if the method of inverse relation (see below) has been able to overcome systematic errors and biases tending to increase the derived value of $H_0$.

The idea that one has to take care of all biases (Malmquist biases of different kind, Lutz and Kelker bias) seems to have been generally accepted. Nevertheless, we find it useful to make here some practical notes.

\subsection{How to correct for the Malmquist bias}
\label{sec2.4}

The Malmquist bias and the cluster incompleteness bias (similar to the Malmquist bias for a sample at a constant distance, like galaxies in a galaxy cluster, or Cepheid stars in an external galaxy) have been modelized \cite{bias}. One can imagine applying the statistical model, in order to correct the bias. However, the correction is model-dependent through the input parameter describing the sources of dispersion. Nevertheless, it is useful at least to visualize the improvement. In the next section, we will use these models in some figures.

Another method, proposed by Schechter \cite{schechter},  has been used by Tully \cite{invtf}. It consist in applying the inverse regression of the linear relation (e.g. the TF relation) instead of the direct relation. Let us explain briefly. When one calculates the linear regression, say $y=a.x + b$, the free parameters $a$ and $b$ are calculated with the direct regression assuming that $x$ has no uncertainty. The inverse regression does the same, but assuming, on the contrary, that all uncertainties are on $x$.  Of course, in real life, both $x$ and $y$ have their own uncertainties. The Malmquist bias results from the fact  that the real sample is truncated on the $y$ axis, as shown in Fig. \ref{spaen}. The inverse regression is not affected by the distorted sample in $y$ (i.e. in magnitude). 
Unfortunately, one cannot correct properly a non-linear effect with a linear one. Furthermore, the inverse relation does not take into account the existing uncertainties on both axes.
Though in principle good for a determination of the Hubble constant, in practice the inverse relation method has its own problems of systematic errors \cite{bias,2kind,invrel}.

 The safest method to correct for the bias for an accurate determination of $H_0$ consists in removing all data beyond the limit of completeness (right side of the dashed vertical line in Fig. \ref{spaen}). This is somewhat similar to the rule of thumb formulated  by Lutz and Kelker. The major drawback of these rules comes from the drastic reduction of the sample. This cannot be accepted in some studies where a large sample is essential. For instance, in the study of the cosmic flow, Tully et al. \cite{laniakea} used the inverse relation, as their main target was not $H_0$, but a detailed cosmography of the local universe.

Increasing the quality of measurements is another important way to reduce bias. Unfortunately, we have no means to reduce the intrinsic dispersion. Any statistical study is prone to be affected by biases.

\section{Bias in the Cepheid Period Luminosity relation }
\label{sec3}

Most secondary distance criteria obey a quasi-linear relationship between the absolute magnitude and one or several observable parameters (e.g., velocity rotation for the TF-relation, Period and colour/metallicity for the PL relation, width of the luminosity curve for the SNIa, etc...). One can naturally ask whether the PL relation may be affected by a bias, similar to the one we discussed in the case of the TF relation. One of us (PT) has pointed out for a long time that such a linear relationship is prone to being affected by biases \cite{pt84}.

\subsection{Why could Cepheids be biased ?  }
\label{sec3.1}
In 1987, one of us (PT) \cite{pt87} identified a new kind of bias, the so-called {\it{population incompleteness bias}}, a Malmquist-type bias, encountered in galaxy clusters. In 1988, Sandage \cite{as88} noticed that truncating a complete sample of Cepheids in the Large Magellanic Cloud (LMC),  leads to too shallow a slope of the PL relation.  This is precisely a population incompleteness bias. In particular, we showed that it affects the HST results \cite{Ref22}.
As we said before the incompleteness of the sample is connected first to the distance, but other parameters may contribute to stimulate the incompleteness (absorption, amplitude of variation, etc.). Hereafter, this bias will be called  "Cepheid Bias", for  sake of simplicity.

During their evolution, stars are moving in the Herztsprung-Russell diagram, a representation of the absolute magnitude {\it{vs.}} effective temperature (or alternatively, colour index or spectral type). At a given stage, when a star has formed enough Helium, it arrives in a region where the so-called $\kappa$-mechanism\footnote{The letter $\kappa$ refers to the absorption by ionized Helium. }
may initiate and maintain a pulsation of the star. Stars will cross this region (the instability strip)  during their evolution. During this passage, the physical conditions in the stellar atmospheres  are modified all over the width of the strip. This can explain an intrinsic dispersion among the pulsating stars, like the Cepheids. All the characteristics of the PL relation can be understood through the physical explanation : the amplitude of the pulsation diminishes for larger and larger wavelengths, the slope of the PL relation increases and its scatter diminishes for larger and larger wavelengths (see Madore \cite{madore1}, p132).

The physical relation governing the pulsation of these stars tells that the period of pulsation is proportional to the inverse of the square root of the mean density. Using classical definitions, one can express the physical relation as an accurate statistical relation with observable parameters. 1) the total luminosity proportional to the square of the radius and the fourth power of the temperature, 2) the relation between mass and luminosity (also dependent on the effective temperature), and 3) the relation between the effective temperature and a colour-index (difference of  two magnitudes in two photometric bands, e.g. V and I). This relation, the Period-Luminosity-Colour (hereafter PLC) relation, is written in logarithmic scale as :
\begin{equation}
M_V = a \ \log P + b \ (V-I)_0 + c ,
\label{plc}
\end{equation}
where $a$, $b$, $c$ are constants and $M_V$ is the absolute magnitude in a photometric band V, $P$ is the period of pulsation (generally expressed in days for normalization purpose), $(V-I)_0$  is the intrinsic colour index expressed, for instance, from photometric bands V and I. It is noteworthy  that, all Cepheid magnitudes being variable, they must be measured at the same phase of the period or defined with the same rule.

 Eq. \ref{plc} is a statistical relation. Its scatter is a first origin of the incompleteness  bias\footnote{The incompleteness has complex origin for Cepheids because to be included in the sample both apparent magnitudes (e.g. V and I)  must be observed during a full phase and this is affected by extinction and amplitude.}. But, there is another fact that could make the bias stronger. 
In order to calculate the intrinsic colour index $(V-I)_0$, we should correct  apparent V and I magnitudes for the interstellar extinction. Unfortunately, this correction is impossible when one uses the classical method. Let us explain.
The absorption by the interstellar medium strongly depends on the wavelength. A star appears redder when seen through a certain amount of interstellar dust located in our Milky Way and in the host galaxy.  This change of colour is called the colour excess. It is the difference between the observed and the intrinsic colour index. The later is precisely the one we are searching for. This means that the calculation of the intrinsic colour excess is impossible from this method without assuming that the intrinsic colour index is constant, or more precisely, that its statistical distribution is the same for the calibrating sample and the studied sample for the concerned extragalactic Cepheids.

The mathematical transcription is as follows. Using V and I apparent magnitudes, the distance modulus, from Eqs. \ref{mu} and \ref{plc}, is:
\begin{equation}
\mu =[ V - R_v ((V-I) - (V-I)_0)] - [a \ \log P + b  (V-I)_0 + c] ,
\label{color}
\end{equation}
where $R_v$ is the ratio of the total-to-the relative absorption, used to correct the apparent $V$ magnitude for the interstellar absorption using the relative change of colour index (colour excess). This can be rewritten as:
\begin{equation}
\mu =W- [a \ \log P + (b-R_v) \ (V-I)_0 + c] ,
\label{color2}
\end{equation}
where $W= V - R_v (V-I)$ is the observable Wesenheit function\cite{bergh}.
Such a presentation transforms the parameter $b$ in $(b-R_v)$\cite{madore2}, but the calculation is exactly equivalent with the classical one (Eq. \ref{color}). $W$ seems more accurate than $V$ because it does not contain the uncertainty due to $(V-I)_0$. This uncertainty goes to the uncertainty on the estimation of $b$.

Anyway, one cannot estimate  the intrinsic colour $(V-I)_0$ (i.e. the effective temperature) with a classical photometric method in order to calculate the distance modulus of a Cepheid from PLC relation\footnote{Note that it would be possible to calculate the colour excess (and thus the intrinsic $(V-I)_0$) if the PLC relation could be replaced by a PL relation. This can be done by writing  two Eq. \ref{color}, in V and I, equating them  and by extracting the colour excess.}. 
This obliges us to assume that, on the mean, both the distant Cepheid sample and the nearby calibrator sample have the same mean intrinsic colour index (e.g. the same $ (V-I)_0 $). The sum of the last two terms in Eq. \ref{color2} is assumed to be a constant (like the second right side term in Eq.\ref{plc}). Thus, the PLC relation  becomes a simple PL relation: 
\begin{equation}
M_V = a \ \log P +  d .
\label{pl}
\end{equation}
In this relation, $d$  is assumed to be constant, but we know that it contains a colour term that may change from a Cepheid to another. This is the second possible cause of a bias.

Madore suggests  a way to bypass the difficulty \cite{madore1}. In short : {\it{Find Cepheids in optical and measure them  in near Infra-red where reddening and scatter are smaller.}}  One cannot exclude that in infra-red, the $b$ coefficient becomes negligible. 
For our test with the PLC relation, we used a new way to bypass the difficulty of calculating $(V-I)_0$  (see section 3.3).

The metallicity of stellar atmospheres may be used as an estimator of the effective temperature, but it may depend also on the initial conditions for the star formation in the host galaxy.  In practice a metallicity correction on  the zero-point ($d$) of the PL relation actually reduces the dispersion.

One can hope that a solution will be found to estimate the effective temperature of a Cepheid (at the phase of the measurement of the apparent magnitude). A spectrum could be used or at least a polychromatic method (see a recent application to LMC of polychromatic method by Inno et al. \cite{inno}).

 How can we check if such a bias exists ? We used several tests and concluded that the bias is real and significant.

\subsection{ Use of the local Hubble law as a reference for distances  }
\label{sec3.2}
\begin{figure}
 \resizebox{0.50\hsize}{!}{ \includegraphics{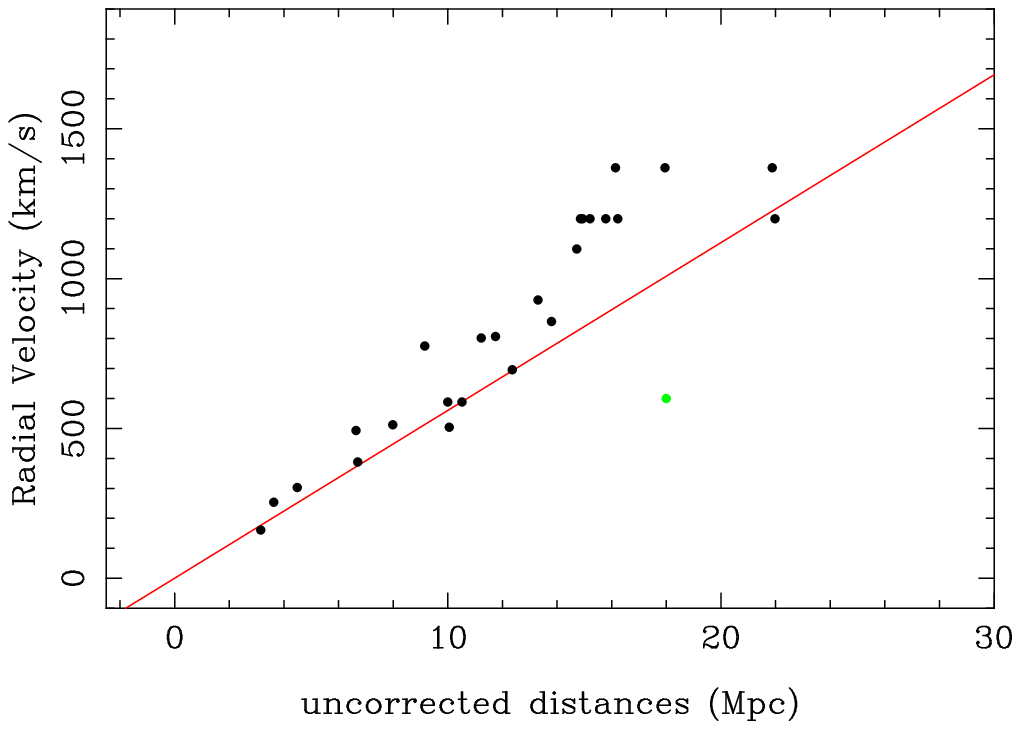}}
 \resizebox{0.50\hsize}{!}{\includegraphics{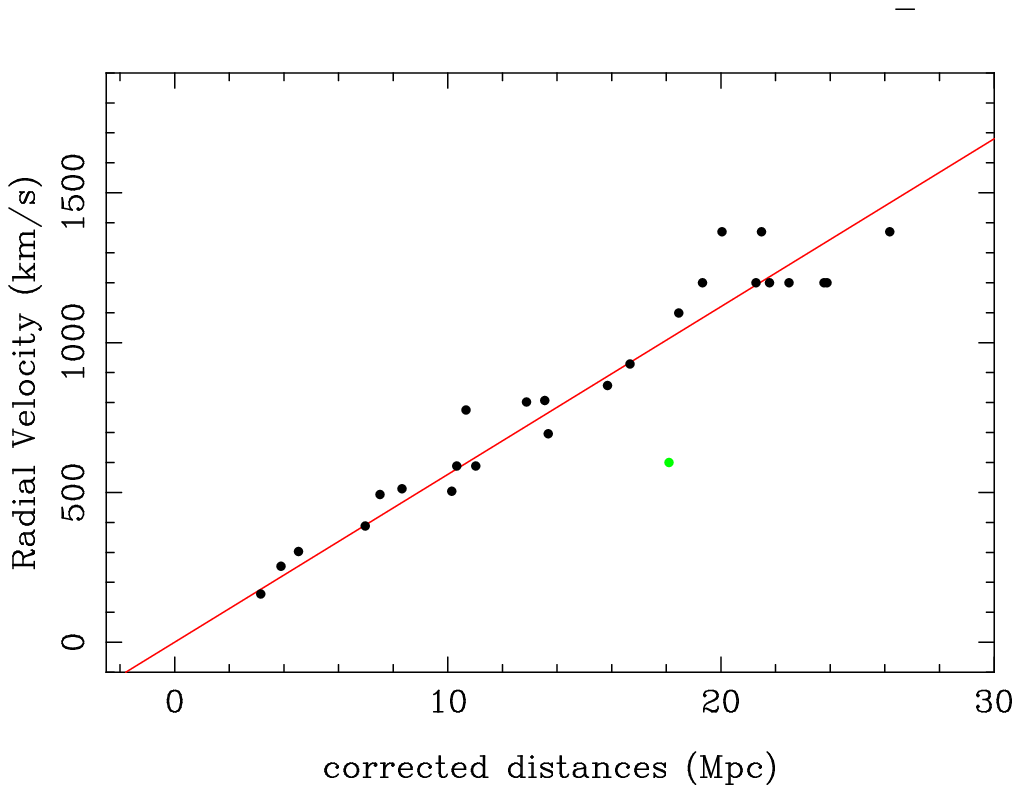}}
\caption{  Hubble diagram from the HST PL relation, for uncorrected (a-left) and corrected (b-right) distance moduli  of extragalactic Cepheids. The coloured dot (NGC4414) is excluded because, at 18 degrees from the Virgo cluster, its corrected velocity is uncertain (see \cite{Ref23}). }
\label{hstpl} 
\end{figure}

The distances from the HSTKP observations confirmed that the Hubble law works at surprisingly small scales as observed in several independent studies \cite{disphl,Ref25} and as already found by Sandage  \cite{Ref26}.   The radial velocities, properly corrected for known velocity flows (the motion of our Galaxy with respect to the Local Group and the motion of the Local Group towards the Virgo cluster) give directly unbiased relative distances without any assumption on $H_0$. 
To test if the HSTKP distances, obtained from the PL relation, are biased, we compare them to the unbiased relative distances.
Fig. \ref{hstpl} (left) shows the classical shape of a bias with a departure from linearity at high distances.
It is possible to go  farther by testing if  this behaviour disappears when the mathematical model  (see \cite{bias}) of the bias is applied to HSTKP distances.   Fig. \ref{hstpl}(right) shows that the linearity has been restored, confirming the suspicion.

\subsection{Use of the Period-Luminosity-Colour relation }
\label{sec3.3}
The PLC relation cannot be used easily because of the unknown absorption that prevents us from determining the intrinsic colour index. Nevertheless, for external galaxies, an important part of the extinction takes place inside our own Galaxy, for which we have good models of the extinction in all directions. Thus, if one assumes that the extinction inside the host galaxy is constant, one can use the PLC relation by correcting only for the local extinction. 

\begin{figure}
 \resizebox{0.50\hsize}{!}{ \includegraphics{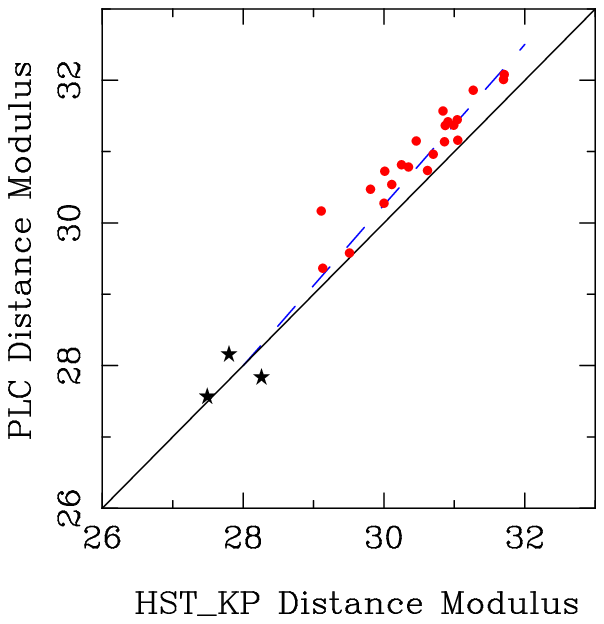}}
 \resizebox{0.50\hsize}{!}{ \includegraphics{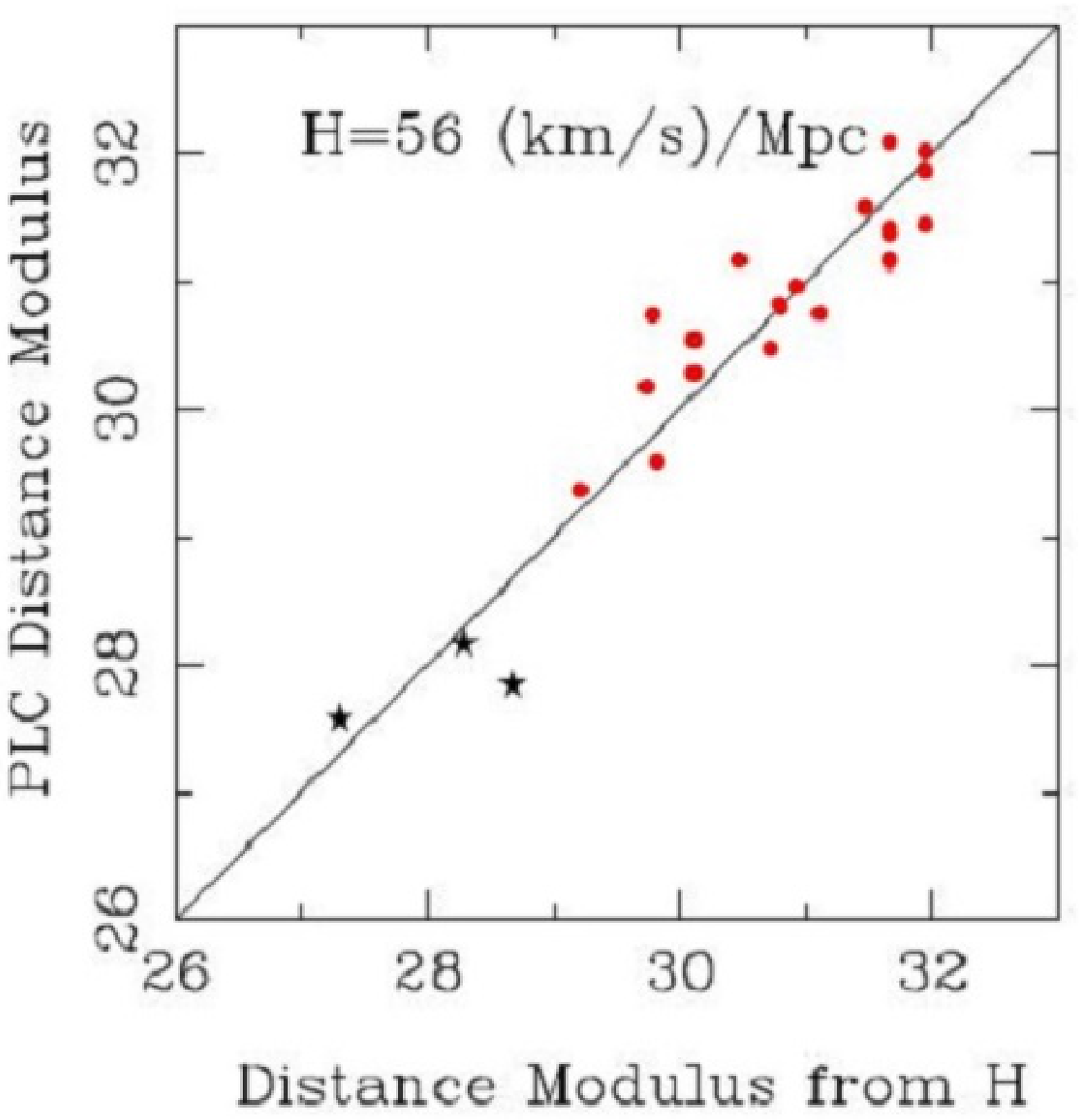}}
\caption{ Comparison of PLC distance moduli (Y-axis) with (on X-axis) : Left : HSTKP distance moduli ;
Right : Unbiased distance moduli from Hubble law with an arbitrary $H_0$ value. For this comparison we used the sample of 26 galaxies given in one paper of the series published by two of us \cite{Ref31}. Data on individual Cepheids are extracted from our Extragalactic Cepheid Database (ECD) \cite{lanoix}. The calibration (black stars) is made with three galaxies classified as {\it{presumably unbiased}} in  \cite{Ref31} for both left and right Figures (see text).}
\label{plc_hst} 
\end{figure}

We calculated  \cite{Ref31} the distance moduli for the HSTKP galaxies. The calibration (black stars in both Fig.\ref{plc_hst}) is made with three galaxies classified as {\it{presumably unbiased}} in  \cite{Ref31}: IC 4182, NGC 3031 and NGC 5253. The plot of the PLC distance moduli {\it{vs.}} HSTKP distance moduli does not follow the first bisect of Fig. \ref{plc_hst}-left,  but shows the typical trend of a bias (biased distances are smaller). In order to be sure that the origin of the discordance does not come from the PLC distances, a comparison is made with unbiased distance moduli derived from velocities through the Hubble law with the value of $H_0$ found in a previous paper (see \cite{Ref31}). The agreement fulfils the hopes (see Fig. \ref{plc_hst}-right).

\begin{figure}[!]
\resizebox{0.55\hsize}{!}{ \includegraphics{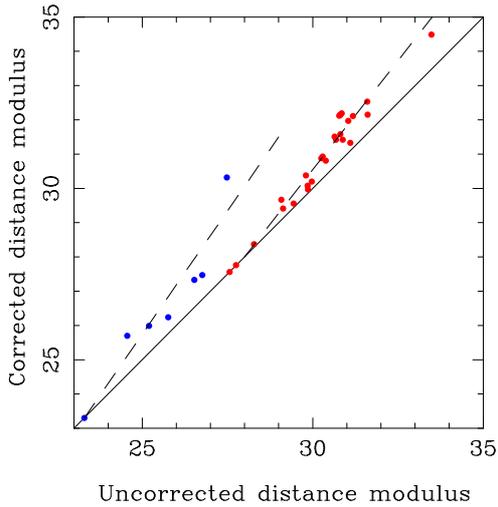}}
\caption{Comparison of corrected  (de-biased) and uncorrected distance moduli of galaxies calculated with a classical PL relation using HST  (red) and old ground-based (blue) observations (taken from ECD).}
\label{ground} 
\end{figure}

One can make an ultimate test by adding ground-based Cepheid observations in a comparison of biased and unbiased distance moduli with two independent calibrations, respectively Galactic Cepheids and LMC.  Fig. \ref{ground} reveals an unexpected, but logical, result. The HST and the ground-based observations  show the same bias trend, but the bias appears at a shorter distance for the ground-based data. 

\section{Questions about the Hubble law}
\label{sec4}
In this part, we will discuss briefly conceptual problems and paradoxes about the Hubble law 
(for more detailed discussions see for instance \cite{tb} and \cite{paradox}). 

\subsection{Different expressions of the Hubble law }
\label{sec4.1}
The historical expression of the Hubble law (Eq. \ref{hubl}) contained a presentation of the observed shift of wavelength as a velocity. Today, the red-shift is measured either as a relative shift in wavelength or, for radio astronomy, as a relative shift in frequency. This means that $z$ can be defined in two ways :
\begin{equation}
z_{\lambda}= \frac {\lambda - \lambda_0}  {\lambda_0} =  \frac {\nu_0 - \nu} {\nu}   \  \  \  \  \    z_{\nu}= \frac {\nu - \nu_0} {\nu_0}=  \frac {\lambda_0 - \lambda}  {\lambda} ,
\label{zis}
\end{equation}
where the "0" index refers to a proper time quantity (i.e. laboratory quantity). The International Astronomical Union (IAU) adopted a recommendation asking radio-astronomers to publish using the optical definition. In  literature the notation $z$ actually refers to $z_{\lambda}$, but it is important to remember that it is a conventional choice.
Both $z_{\lambda}$ and $z_{\nu}$ may be used. 
The transformation $z_{\lambda} \iff z_{\nu}$  results simply from  Eq. \ref {zis}, and $\lambda \nu=c$ and $\lambda_0 \nu_0=c$ 

\begin{equation}
z_{\lambda}= - \frac {z_\nu} {1 + z_{\nu}}   \  \  \  \  \  z_{\nu}= - \frac {z_{\lambda}} {1+ z_{\lambda}} 
\label{zl2zn}
\end{equation}

 With this convention, the published velocities  are all expressed in the $c z_{\lambda}$  system (not in the  $-c z_{\nu}$) and can be compared together. Nevertheless,  $z_{\lambda}$ and $z_{\nu}$ behave differently when distances tend to infinity : $z_{\lambda} \to \infty$ and  $z_{\nu} \to - 1$.

Before discussing  different expressions of the Hubble law, we must emphasize that the luminosity distance $r$, used in practical  cosmology, becomes uncertain at large distances, because some technical corrections on magnitudes (the K-correction and the evolution correction) are less and less reliable when the distance grows, typically when $z_\lambda > 1.$. The luminosity distance is the distance measured assuming an Euclidean universe. In this section, the luminosity distance must be considered as a function of the relevant parameters ($z$, $q_0$...) for the chosen paradigm. Three paradigms are given by Harrison\cite{harrison1}. In all cases,the Hubble law has the form of Eq.\ref{hubl}, but the meaning of each quantity may differ:
\begin{enumerate}
\item The redshift-distance law. The observable shift  in both wavelength and frequency is interpreted as a new effect (the “tired light” hypothesis) that does not result from a true motion. The simplest expression of the Hubble law is then :
\begin{equation}
{c}z={H_0} r 
\label{hl1a}
\end{equation}
For example, one can relate to this paradigm the case of an {\it{a priori}} relation of the redshift $z$ with the proper time, justified by the consequences deduced from this arbitrary choice.  For instance the Hubble law can be written in a static space as: $z_{\nu}=H_0 t$ with some interesting predictions (see e.g. \cite{yhs}).

\item The velocity-distance. If one assumes that the red-shift is due to a Doppler effect (omitting the gravitational red-shift), then the same velocity is found, either with $z_{\lambda}$ or $z_{\nu}$, using the relativistic expressions of the Doppler effect that can be combined into a single expression as :
\begin{equation}
V=c  \mid \frac{A^2-1}{A^2+1} \mid,
\label{vdop}
\end{equation}
with $A$ being either $1+z_{\lambda}$ or $1+z_{\nu}$ and V being positive for a red shift and negative for a blue shift. 

\item The expanding space.  If one considers that the redshift is explained as in the Friedman model, the space-expansion velocity is calculated using the expression of $r$ as a function of the co-moving distance $\chi$ and of the  time varying scale factor $S(t)=r/\chi$:
\begin{equation}
 V = \frac{dr}{dt}= \chi  \frac{dS}{dt} =  \frac{\dot{S}}{S} r = H(t)r
\label{friedmann}
\end{equation}
\end{enumerate}

Except for the second paradigm for which the velocity has a classical meaning limiting it to $c$, the velocity-like parameters of paradigms 1 and 3 can be larger than the velocity of light for distances larger than the Hubble distance $R_H=c/H0$ as noted by Harrison \cite{harrison1}.
Obviously, all expressions are equivalent to Eq. \ref{hubl} at small red-shifts. In fact, to choose the correct expression of the Hubble law, one should know the physical nature of the red-shift. 

\subsection{ The physical nature of the red-shift }
\label{sec4.2}
The first suggestion for a possible cosmological red-shift was the global gravitational red-shift in the static de Sitter universe \cite{ds1,ds2,ds3}.  In the early history of the relativistic cosmology, de Sitter, Eddington \cite{Ref45} and Tolman \cite{Ref46} discussed the possibility to observe this de Sitter effect in a static cosmological model. De Sitter found a static solution of Einstein's equations for an empty universe with the cosmological constant.
	Eddington emphasized that : "{\it{In De Sitter's theory... there is the general displacement of spectral lines to the red in distant objects due to the slowing down of atomic vibrations which... would be erroneously interpreted as a motion of recession}}". In fact, in his famous study Hubble \cite{edwin} refers to the de Sitter effect as a possible explanation of the distance-red-shift law.
Sandage  \cite{alan1} made a list of the most urgent astronomical problems.  The question : "{\it{Is the expansion real?}}"  is in the first place in the cosmological list.
In 1935, Tolman proposed the surface brightness test  to answer this question. Indeed, the ratio of the luminosity to the surface is constant for a uniform source of light in static space. In an expanding universe it must vary as $(1+z)^{-n}$, with $n=4$. In the near infrared, where the extinction is small, Sandage \cite{sbt} found $n$ within the range $2.8$ to $3.5$. This seems to be in favour of a real expansion, and anyway, rules out the static and tired-light model (see below).
Sandage proposed also to test, for the future, \cite{alan2} if the distance between galaxies really increases with time or if the red-shift of a given object (galaxy, quasar) changes with time. In expanding space, $z$ is expected to change according to the relation: 
\begin{equation}
dz/dt = (1+z) \ H_0 - H(z) .
\end{equation}
In terms of radial velocity, the predicted change is very small $dv/dt \approx 1$ (cm/s)/yr, but may be within the reach of future 39-meter Extremely Large telescope (ELT) \cite{elt}.

In the frame of the standard $\Lambda$CDM Big Bang model, the cosmological red-shift is interpreted as space expansion, not as a Doppler effect. Mathematically, the space expansion is the time dependence of the distance between galaxies: $r(t)=S(t)\chi$, where $\chi$ is the co-moving (constant) coordinate. In the Friedmann model the exact linear velocity – distance relation is a strict mathematical consequence of the homogeneity of space. So, why is the Hubble law valid in the nearby space, where the matter is not homogeneously distributed?
One of us \cite{fract} noted that the global gravitational red-shift within the fractal matter distribution with fractal dimension $D = 2$ can produce a linear Hubble law. A cosmological model based on this idea was developed in \cite{fractmod}.
A review of tests on the nature of the expansion is given in \cite{lopez}.
\subsection{At what level on the scale does the Hubble law start ?  }
\label{sec4.3}
The linear Hubble law is clearly established around the Local Group of galaxies in the interval of scales $1$ to $100$ Mpc \cite{linhl}. The distribution of matter is not uniform at that scale and is characterized by fractal power-law behaviour \cite{klun,sylos}.
Two principal questions arise here: 1) Why is the linear $z$ - $r$ relation observed, while according to the standard cosmology, the Hubble law is the strict consequence of the homogeneity? \cite{teka,yuri2}  2) And why is the dispersion of the Hubble law so small \cite{disphl}?

	Can we imagine that the Hubble law works also at a very small scales, inside our Galaxy or inside the solar system, or even at still smaller scales? Is it simply hidden by stronger interactions? This could be similar with the gravitation which is hidden between ordinary objects. If it works at all scales, would it be possible to detect the Hubble law without a scale invariant reference?

It would be possible to detect a local expansion by comparing two clocks, one based on a length (quartz or  "whispering gallery" (WG)) and another based on a reference frequency (atomic clock), assuming it to be scale invariant. Both comparisons were made (quartz, and WG), showing first a relative drift of  the order of magnitude of the Hubble constant. New measurements showed a much smaller drift\cite{biz}. We must conclude that a local expansion is not observed at extremely small scale dominated by local matter and we follow Landau \& Lifschitz \cite{ll} who wrote: {\it{the conclusion that the bodies are running away with increasing [scale factor] $a(t)$ can only be made if the energy of interaction of the matter is small compared to the kinetic energy of its motion in recession}}.

	Recently, a new test of the global space expansion, to be performed within the Solar system, was proposed \cite{kope,kope2}. Although the solar system, the planets, and the galaxies do not expand, the free photons within these objects will feel the global space expansion of the universe. The predicted blue shift of such photons is about $d \nu / \nu = H \ t$.


\section{Conclusions  }
\label{sec5}
\subsection{Biases }
\label{sec5.1}
At the demand of a referee, we will express our point of view.
A first conclusion is that the attempts to use the observations at their extreme limit are generally hampered by hidden selection effects that appear as a surprise, but might have been anticipated in the light of the history of cosmic distance measurements. 
We claimed for many years that biases are always present in measurement of distances. This is due to the fact that it is very difficult to construct a distant sample having the same constitution as the calibrating one. For instance, consider Fig.\ref{ground} and imagine that we had calibrated the HST distances with the ground-based distances. We would have had the impression of a linear Hubble law extending over a huge range of distances. 
On the basis of our studies, we believe that the $H_0$ is probably  less than 70 by several units. We hope that the results of future observations (GAIA, JWST, ELT39m, etc...) will clarify the fuzzy word of "several".
To reduce the effects of biases we must reduce the dispersion by selecting very homogeneous samples and, most importantly, we must systematically study the completeness of them, in order to reject  all objects beyond the completeness limit. Remember the rule of thumb of Lutz and Kelker. This is the only way to obtain a reliable $H_0$ by classical measurements in the galaxy universe.

\subsection{Important facts }
\label{sec5.2}
Four facts  play a major role in cosmology: 
\begin{itemize}
\item The discovery of the redshift correlated with distances, {\bf{the Hubble law}},  is, of course, the major fact, even if its profound nature is not fully understood so far.

\item In 1937, Zwicky discovered strange evidence  \cite{zwicky} : the mass of a galaxy cluster calculated by a dynamical study differs from the mass inferred from the direct count of visible galaxies. This is the origin of  {\bf {Dark Matter}}, later confirmed from galaxy rotation curves and gravitational lensing. The SCM interprets this  by introducing a very large amount of non-observable exotic matter (cold non-baryonic particles). 
An empirical model (MOND  \cite{mond}), which modifies the Newtonian dynamics at small accelerations ($a \approx H_0 c$) can explain observations without introducing non-baryonic Dark Matter. It could be just a fortuitous coincidence, but some people consider the model (and its relativistic formulation \cite{beken}) with some interest \cite{blanchet}.

\item The observational fact of the high level of isotropy of the 3K cosmic microwave background radiation means that causally unconnected regions of the universe have the same physical parameters,  and that the initial Hubble constant in the moment of creation must be fine-tuned to produce the observed  {{flat universe}}  with nearly the critical density. The  {\bf {Inflation}} at the beginning of the universe was invented by Guth \cite{guth} to overcome these problems. 

\item The discovery  \cite{perl,riess} of the observed extra faintness of distant SN Ia supernovae at high red-shifts, as compared with the brightness expected from the Friedmann model containing just gravitating matter. Within the standard cosmological model, this corresponds to the acceleration of the expansion, and leads to the {\bf{Dark Energy}}, formulated by re-using the cosmological constant invented by Einstein to make a static universe possible.
\end{itemize}

\subsection{A search for a common link }
\label{sec5.3}
 All these ingredients could be like Ptolemaic epicycles. Would it be possible to explain them through a single hypothesis? They have something in common: They all appear as a kind of unexplained matter-energy that has something to do with the Hubble law. What is the source of this matter-energy? 

The vacuum contains energy. Does it contribute or explain the missing energy? Time is linked with energy. The conservation of the energy is deduced from the uniformity of time (Noether's theorem). Is this a clue to find a common explanation to these new ingredients introduced in cosmology?
Furthermore, it is admitted that in General Relativity the  fundamental law of conservation of energy is not valid globally but only as a local concept (see Peebles\cite{peebles} \S6, p139). Let us explain the origin of this assertion (for more details, read \S101 in \cite{ll}). 
Without gravitational field, the expression of  the conservation of the energy-momentum tensor $T_{ij}$  is given by the divergence ${\partial}_j  T^{i  j}=0 $. For gravitation,  it must be written with covariant derivatives  : ${\cal{D}}_j  T^{i  j}=0 $. The calculation gives a sum of two terms that does not express any general conservation law. Nevertheless, with a convenient choice of the coordinate system (the choice is permitted for tensors), the second term vanishes in one point, leading to the classical conservation laws at this point. 
It is then possible to calculate how to transform $T_{ij}$ at this point in such a way that the covariant divergence is identical with the normal divergence. It has been found that $T_{ij}$ must be replaced by $T_{ij}+ t_{ij}$, where $t_{ij}$ is the energy of the gravitational field, not expressed in $T_{ij}$. Unfortunately, the conservation law cannot be applied to any reference system, because  $t_{ij}$  is  not a tensor but a pseudo-tensor. Nevertheless, by imposing some physical conditions (an isolated system of masses in an asymptotically flat space), it is possible to define a class of reference systems (a "world-canal") for which the conservation laws work (including the energy conservation law). 
Is it possible to have an accurate global description of cosmology in the light of recent discoveries without restoring the global status of the energy conservation?

\medskip
We conclude with a comment on fortuitous coincidences. In previous subsections, we encountered some numerical coincidences that can be simple numerical haphazard combinations. Another one has been analysed by outstanding scientists (Eddington, Weyl, Dirac). By combining some fundamental quantities, it is possible to build two dimensionless large numbers. It seems possible to assume that these numbers are equal because of their magnitude (about $10^{40}$). The intriguing problem is that the Hubble constant appears in the expression of one of these large numbers.  This looks very uncertain, but we must keep in mind, that sometimes this sort of coincidence may reveal something profound. We have two famous examples :1) In 1950, Herzberg wrote : " {\it {From the intensity ratio of the lines with K=0 and K=1 a rotational temperature of 2.3°K follows, which has of course only a very restricted meaning.}}". This was in fact the Cosmic Microwave Background Radiation discovered 15 years later by Penzias and Wilson.   2) In 1857, Kirchoff did not consider seriously the numerical coincidence $\epsilon_0.\mu_0 =c^{-2}$  obtained by Weber and Kohlrausch. Maxwell gave a justification of this expression 7 years later.

One cannot build a theory on such coincidences, but this may help us to find a way from known precepts to new ones. Heisenberg wrote in his "1942 Manuscript" : "{\it {The formal deduction is powerless to throw a bridge on the abyss [between the system of known concepts and the system of new concepts]}}"

\begin{acknowledgements}
Yu. B. thanks for support from the Saint-Petersburg State University research project No.6.38.18.2014. We are very grateful to our collaborators over many years Lucienne Gouguenheim, Gilles Theureau, Jean-No{\"{e}}l Terry, Chantal Petit and Mikko Hanski, and remember with admiration the late Lucette Bottinelli and Timo Ekholm. We want to sincerely thank referees for their valuable contributions.
\end{acknowledgements}


\begin{thebibliography}{}
\bibitem{vesto}
Slipher, V.M., On spectrographic observations of nebulae and clusters, PAAS. 4, 284 (1922)
\bibitem{lemaitre}
Lema\^{\i}tre, G., Un univers homog\`{e}ne de masse constante et de rayon croissant rendant compte de la vitesse radiale des n\'{e}buleuses extragalactiques, AASB. 47, 49 (1927)
\bibitem{lundmark}
Lundmark, K., The motions and distances of spiral nebulae, MNRAS, 85, 865 (1925)
\bibitem{edwin}
Hubble, E.,  A relation between distance and radial velocity among extragalactic nebulae. Proc. Nat. Acad. Sci. 15, 168-173 (1929)
\bibitem{deba1}
Maddox, J.,  Dispute over scale of Universe. Nature 307, 313 (1984)
\bibitem{deba2}
Giovanelli, R.,  Less expansion more agreement. Nature 400, 111-112 (1999)
\bibitem{Ref6}
Roberts, M.S., The neutral Hydrogen content of late-type spiral galaxies, Astron. J., 67, 437 (1962)
\bibitem{Ref7}
Gouguenheim, L., Neutral Hydrogen content of small galaxies, Astron. Astrophys. 3, 281 (1969)
\bibitem{tf}
Tully, R.B., Fisher, R., A new method for determining distances to galaxies, Astron. Astrophys. 54, 661 (1977)
\bibitem{bias}
Teerikorpi, P.,  Observational selection bias affecting the determination of the extragalactic distance scale. Ann.Rev.Astron.Astrophys. 35, 101-136 (1997)
\bibitem{2kind}
Teerikorpi, P.,  The inverse Tully-Fisher relation, Astro Lett. and Comm., 31, 263 (1995)
\bibitem{jnt}
Terry, J.N.,  Paturel, G., Ekholm, T., Local velocity field from sosie galaxies : The Peeble's model, Astron. Astrophys.,393, 57 (2002) 
\bibitem{spaenh}
 Spaenhauer , A.M., A systematic comparison of four methods to derive stellar space densities. Astron. Astrophys. 65,313 (1978)
\bibitem{Ref14}
Bottinelli L., Gouguenheim, L., Paturel, G., Teerikorpi, P., The Malmquist bias and the value of $H_0$ from the Tully-Fisher Relation, Astron. Astrophys. 156, 157 (1986)
\bibitem{Ref15}
Bottinelli L., Gouguenheim, L., Paturel, G., Teerikorpi, P., The Malmquist bias in the extragalactic distance scale : Controversies and misconceptions, Astrophys. J. 328, 4, (1988)
\bibitem{lutz}
Lutz T.E., Kelker D.H., On the Use of Trigonometric Parallaxes for the Calibration of Luminosity Systems: Theory, PASP 85, 573 (1973)
\bibitem{feast}
Feast, M.W., Catchpole, R.M., The Cepheid period-luminosity zero-point from HIPPARCOS trigonometrical parallaxes, MNRAS 286, L1-L5 (1997)
\bibitem{Ref19}
Freedman, W.L. et al., Final results from the Hubble Space Telescope Key Project to measure the Hubble constant. Astrophys.J 553, 47-72
\bibitem{beaton}
Beaton, R.L., Freedman, W.L., Madore, B.F, et al., The Carnegie-Chicago Hubble Program I : A new approach to the distance ladder, Astrophys. J. 832,2101 (2016)
\bibitem{pt84}
Teerikorpi, P., Malmquist bias in a relation of the form $M=a\ P+b$, Astron. Astrophys., 141, 407 (1984)
\bibitem{hamuy}
Hamuy, M., Phillips, M.M., Suntzeff, N.B., et al, The Hubble diagram of the Calan/Tololo Type Ia Supernovae and the value of $H_0$, Astron. J., 112, 2398 (1996)
\bibitem{riess}
 Riess, A.G., Filippenko, A.V., Challis, P. et al., Observational evidence from supernovae for an acccelerating universe and a  cosmological constant, Astron. J., 116, 1009 (1998).
\bibitem{perl}
 Perlmutter, S., Aldering, G., Goldhaber, G., et al., Measurement of $\Omega$ and $\Lambda$ from 42 high-red-shift supernovae. Astrophys J. 517, 565 (1999)
\bibitem{wmap}
Bennett, C.L., Larson, D., Weiland, J.L., et al., Nine-Year Wilkinson Microwave Anisotropy Probe (WMAP) observations : Final Maps and Results, Astrophys. J. Supl., 208, 20 (2013)
\bibitem{planck}
Planck Collaboration: Ade, P.A.R., Aghanin N., Arnaud, M., et al, Planck 2015 results. XIII. Cosmological parameters, Astron. Astrophys., 594, 15 (2016)
\bibitem{gieren}
Gieren, W., Fouqu\'{e}, P., Gomez, M., Cepheid Period Radius and Period Luminosity Relations and the Distance to the Large Magellanic Cloud, Astrophys. J., 496, 17 (1998)
\bibitem{benedict}
Benedict, G.F., Mc Arthur, B.E., Feast, M.W., et. al., Hubble Space Telescope fine guidance sensor parallaxes of Galactic Cepheid variable stars: Period Luminosity Relation, Astron. J.,133,1810 (2007)
\bibitem{4258}
Herrnstein, J.R.,Moran, J.M., Greenhill, L.J. et al, A geometric distance to the galaxy NGC 4258 from orbital motions in a nuclear gas disk, Nature 400, 539 (1999)
\bibitem{hoffmann}
Hoffman, S.L., Riess, A.G., Macri, L.M., Hoffmann,  et al., Optical Identification of Cepheids in 19 Host Galaxies of Type Ia Supernovae and NGC 4258 with the Hubble Space Telescope, Astrophys. J., 830, 10  (2016)
\bibitem{riess2}
Riess, A.G., Macri, L.M., Hoffman, S.L.  et al., A 2.4\% Determination of the Local Value of the Hubble Constant , Astrophys. J., 826, 56  (2016)
\bibitem{laniakea}
Tully, R.B., Courtois, H.M., Dolphin, A.E., et al, Cosmicflow-2 : Data, Astron. J., 146, 86 (2013)
\bibitem{pt87}
Teerikorpi, P., Cluster population incompleteness and distances from the TF relation - Theory and numerical example,  Astron. Astrophys., 173, 39 (1987)
\bibitem{as88}
Sandage, A., Cepheids as distance indicators when used near their detection limit, PASP, 100, 935 (1988)
\bibitem{schechter}
Schechter, P.L., Mass-to-light ratios for Elliptical galaxies, Astron.J., 85, 801 (1980)
\bibitem{invtf}
Tully, R.B., Origin of the Hubble constant controversy,  Nature 334, 209 (1988)
\bibitem{invrel}
Teerikorpi, P., Ekholm, T., Hanski, M.O., Theureau, G. Theoretical aspects of the inverse Tully-Fisher relation as a distance indicator: incompleteness in $logV_{max}$, the relevant slope, and the calibrator sample bias. Astron. Astrophys. 343, 713 (1999)
\bibitem{Ref22}
Teerikorpi, P., Paturel, G. Evidence for the extragalactic Cepheid distance bias from the kinematical distance scale. Astron.Astrophys. 381, L37-L40 (2002)
\bibitem{madore1}
Madore, B.F., in "From the Realm of the Nebulae to Populations of Galaxies", Eds. D'Onofrio, M., Rampazzo, R., Zaggia, S., ed. Springer, page 132, (2016)
\bibitem{madore2}
Madore, B.F., The Period Luminosity Relation : IV - Intrinsic Relation and Reddenings for the Large Magellanic Cloud Cepheids, Astrophys. J. 253, 575 (1982)
\bibitem{bergh}
Van den Bergh, S., The galaxies of the Local Group, JRASof Canada, 62, 145 (1968)
\bibitem{inno}
Inno, L., Bono, G., Matsunaga, N., The panchromatic view of the Magellanic Cloud from classical Cepheids I: Distance, Reddening and Geometry, Astrophys. J. 832, 176 (2016)
\bibitem{Ref23}
Ekholm, T., Lanoix, P., Teerikorpi, et al., Investigation of the Local Supercluster velocity field Astron.Astrophys. 351, 827-833 (1999)   
\bibitem{disphl}
Ekholm, T., Baryshev, Yu., Teerikorpi, et al.. On the quiescence of the Hubble flow in the vicinity of the Local Group: A study using galaxies with distances from the Cepheid PL-relation. Astron.Astrophys. 368, L17-L20 (2001)
\bibitem{Ref25}
Karachentsev, I.D. et al. The very local Hubble flow. Astron. Astrophys. 389, 812-824 (2002)
\bibitem{Ref26}
Sandage, A., The red-shift-distance relation. IX Astrophys. J. 307, 1  (1986)
\bibitem{Ref31}
Paturel, G., Teerikorpi, P., The extragalactic Cepheid bias: a new test using the Period-Luminosity-color relation, Astron. Astrophys 452, 423-430 (2006)
\bibitem{lanoix}
Lanoix, P., Garnier, R., Paturel, G., et. al., Extragalactic Cepheid Database, Astron. Nachr. 320, 1 (1999)
\bibitem{tb}
Baryshev, Yu. \& Teerikorpi, P., 2012, Fundamental Questions of Practical Cosmology (Springer, Berlin)
\bibitem{paradox}
 Baryshev, Yu, V., Paradoxes of the cosmological physics in the beginning of the 21-th  century In : Particle and Astroparticule Physics, gravitation and cosmology : predictions, observations and new projects, pp297-307 (2015) arXiv: 1501-01919
\bibitem{harrison1}
 Harrison, E.R.: The red-shift-distance and velocity-distance laws. Astrophys. J. 403, 28 (1993)
\bibitem{yhs}
Sanejouand, Y.H., A simple Hubble like law in lieu of dark energy, arXiv : 1401.2919v6 (2015)
\bibitem{ds1}
 de Sitter W.,  On Einstein's theory of gravitation and its astronomical consequences I, MNRAS, LXXVI. 9, 699 (1916)
\bibitem{ds2}
 de Sitter W., On Einstein's theory of gravitation and its astronomical consequences  II, MNRAS LXXVII. 2, 155 (1917)
\bibitem{ds3}
de Sitter W., On Einstein's theory of gravitation  and its astronomical consequences III, MNRAS LXXVIII. 1, 3 (1917)
\bibitem{Ref45}
 Eddington, A.S., 1923, The mathematical theory of relativity, (Cambridge), p.161
\bibitem{Ref46}
Tolman, R.C., O, the astronomical implications of the de Sitter line element of the universe, Astrophys. J. 69, 245 (1929) 1929
\bibitem{Ref47}
Sandage, A., 1975, in Galaxies and the Universe (The University of Chicago Press 1975)
\bibitem{alan1}
Sandage A.,  Astronomical problems for the next three decades. In Key Problems in Astronomy and Astrophysics, Mamaso A. and Munch G. eds., Cambridge University Press (1995)
\bibitem{sbt}
Sandage A., The Tolman surface brightness test for the reality of the expansion, V. Provenance of the test and a new representation of the data for three remote Hubble space telescope galaxy clusters. Astron. J. 139, 728 (2010)
\bibitem{alan2}
 Sandage, A., The change of red-shift and apparent luminosity of galaxies due to the deceleration of the expanding universes. ApJ 136, 319 (1962)
\bibitem{elt}
 Liske, J., Grazian, A., Vanzella, E., et al., Cosmic dynamics in the era of extremely large telescopes, Mon. Not. R. Astron. Soc. 386, 1192 (2008)
\bibitem{fract}
Baryshev, Yu.V., The hierarchical structure of metagalaxy – a review of problems, Reports of Special Astrophysical Observatory of the Russian Academy of Sciences 14, 24 (1981) (English translation: 1984 Allerton Press)
\bibitem{fractmod}
 Baryshev, Yu. V. Field fractal cosmological model as an example of practical  cosmology approach, in ”Practical Cosmology”, Proceedings of the International Conference held at Russian Geographical Society, 23-27 June, 2008, Vol.2, p.60 (2008), (arXiv:0810.0162)
\bibitem{lopez}
Lopez-Corredoira, M., Tests of the expansion of the Universe, arXiv:1501.01487 (2015)
\bibitem{linhl}
 Sandage A., Reindl B. and Tammann G., The Linearity of the Cosmic Expansion  Field from 300 to 30,000 km s-1 and the Bulk Motion of the Local Supercluster with Respect to the Cosmic Microwave Background, Astrophys. J.,714, 1441 (2010)
\bibitem{klun}
Teerikorpi, P., Hanski, M., Theureau, G., et al.: The radial space distribution of KLUN-galaxies up to 200 Mpc: Incompleteness or evidence for the behavior predicted by fractal dimension ≈2?, Astron. Astrophys. 334, 395 (1998)
\bibitem{sylos}
Sylos Labini F., Inhomogeneous Universe, Class. Quant. Grav. 28 , 4003  (2011).
\bibitem{teka}
Tekhanovich, D.I., Baryshev, Yu.V., Global Structure of the Local Universe according to 2MRS Survey; ISSN 1990-3413, Astrophysical Bulletin 71, No. 2, 155-164, Ed. Pleiades Publishing, (2016)  (arXiv: 1610.05206)
\bibitem{yuri2}
Y.V. Baryshev, Two fundamental cosmological laws of the Local Universe, Proceedings of the International Conference, Cosmology On Small Scales, Local Hubble Expansion and Selected Controversies in Cosmology, Prague, September 21–24, 2016, Edited by Michal Kr\v{i}zek and Yurii V. Dumin, Institute of Mathematics, Czech Academy of Sciences, Prague, pp.9 – 22, (2016) ( arXiv:1610.05943)
\bibitem{biz}
Wiens, E., Nevsky, A.Yu., Schiller, S., Resonator with ultra-high stability as a probe for Equivalence-Principle-violating physics, arXiv: 1612.01467V1, (2016)
\bibitem{ll}
Landau, L.D, Lifschitz, E.M., Theory of Fields, Mir Ed., Moscow (1970)
\bibitem{kope}
Kopeikin S., Celestial Ephemerides in an Expanding Universe, Phys. Rev. D, 86, 064004 (2012)
\bibitem{kope2}
 Kopeikin S., Local gravitational physics of the Hubble expansion, Eur. Phys. J. Plus (2015) 130, 11 (2015) (arXiv:1407.6667)
\bibitem{zwicky}
Zwicky, F., On the masses of nebulae and of clusters of nebulae, Astrophys. J. 86, 217 (1937)
\bibitem{mond}
Milgrom, M.  A modification of the Newtonian dynamics as a possible alternative to the hidden mass hypothesis, Astrophys. J. 270, 365 (1983)    MOND
\bibitem{beken}
Bekenstein, J.,Relativistic gravitation theory for the modified Newtonian dynamics paradigm, Physical Rev.D70, 083509, 2004
\bibitem{blanchet}
Blanchet, L., Gravitational polarization and the phenomenology of MOND, Class. Quant. Grav. 24, 3529 (2007)
\bibitem{guth}
 Guth, A.H., Inflationary universe : A possible solution to the horizon and flatness problem. Phys. Rev. D23, 347 (1981)
\bibitem{peebles}
Peebles; P.J.E, Principles of Physical Cosmology.Princeton University Press, Princeton (1993) (\S6, page139)






\end{thebibliography}
\end{document}